\documentclass[prl, showpacs, twocolumn, floatfix]{revtex4}

\usepackage{graphicx}
\usepackage{amsmath, amsfonts, amssymb, bm}
\usepackage[]{psfrag}

\psfragscanoff
\begin{document}

\title{ Dynamic two-center resonant photoionization in slow atomic collisions. } 

\author{ A. B. Voitkiv }
\author{ C. M{\"u}ller }
\affiliation{ Institute for Theoretical Physics I, Heinrich-Heine-Universit\"at D\"usseldorf, Universit\"atsstrasse 1, 40225 D\"usseldorf, Germany }
\author{S. F. Zhang}
\author{X. Ma}
\affiliation{Institute of Modern Physics, Chinese Academy of Sciences, Lanzhou 730000, China} 

\date{\today}

\begin{abstract}
An additional channel opens for photoionization of atom $A$ 
by an electromagnetic field if it traverses a gas of atoms $B$  
resonantly coupled to this field. We show that 
this channel, in which $A$ is ionized via resonant 
photoexcitation of $B$ with subsequent energy transfer 
to $A$ through two-center electron correlations 
and which is very efficient when $A$ and $B$ constitute 
a bound system, can  
strongly dominate the ionization of $A$ also in collisions 
where the average distance between $A$ and $B$ exceeds 
the typical size of a bound system by orders of magnitude. 
 
\end{abstract}
 

\maketitle

\section{Introduction} 

The breakup of bound microscopic systems by photoabsorption 
is characterized by well defined energy and momentum 
transfers, which often enables one to extract precise information  
about the process and the system itself. 
Studies of photo-induced breakup reactions
-- such as atomic photoionization (PI), molecular photodissociation, 
and nuclear photo-disintegration -- 
have therefore played a key role in our understanding 
of the structure and dynamics of matter on a microscopic scale. 

Electron correlations are omnipresent in the quantum world, 
ranging from atoms and small molecules to organic macromolecules and solids.  
They drive autoionization of atoms and ions and 
mutual electron transitions in high-energy ion-atom collisions 
\cite{mcguire-book}-\cite{abv-book-2008}, 
can result in de-excitation reactions 
in very slow atomic collisions \cite{Smirnov} 
and in ultracold quantum gases \cite{Weidemueller}, 
govern energy transfer between chromophores \cite{Forster} 
and lattice dynamics in polymers \cite{Suhai},  
and are even responsible for magnetism and superconductivity \cite{solids}. 
Electron correlations coupling different atoms, 
which occur in bound systems with more than one 
atomic center, lead to inter-atomic Coulombic decay 
of inner-shell vacancies \cite{icd-all} 
-- an autoionization-type reaction representing 
a kind of a two-center Auger decay --    
observed in dimers and clusters \cite{clusters,resonantICD,He} 
and water molecules \cite{ICDexpH2O}. 
%Autoionization reactions in bound systems with 
%more than one atomic center driven 
%by inter-atomic electron correlations 
%lead to the rich physics of inter-atomic Coulombic decay \cite{icd-all} 
%observed in dimers and clusters \cite{clusters,resonantICD,He} 
%and water molecules \cite{ICDexpH2O}. 
%Autoionization reactions in bound systems with 
%more than one atomic center driven 
%by inter-atomic electron correlations 
%are responsible for the rich phenomenology 
%of inter-atomic Coulombic decay  \cite{icd-all} 
%observed in dimers and clusters \cite{clusters,resonantICD,He} 
%and water molecules \cite{ICDexpH2O}.
Inter-atomic electron correlations can 
greatly enhance recombination processes 
\cite{2CDR,our-new-paper} and lead to resonances 
in electron scattering on two-atomic systems \cite{our-new-paper}.  

Particularly clean manifestations 
of electron correlations are revealed in some PI processes, 
e.g. in single-photon double ionization \cite{SPDI},
in atomic autoionization triggered by photoabsorption \cite{Rzazewski},
in non-sequential double ionization in strong laser fields \cite{NSDI} 
and resonant two-center PI \cite{we-2010}. 
In the latter, ionization of a large-size molecule 
occurs via resonant photoabsorption by one of its atoms 
with subsequent transfer of excitation energy 
via two-center electron correlations to another atom 
leading to its ionization. 
This two-center ionization channel can be remarkably effective 
strongly dominating over the usual direct single-center PI and  
it was experimentally observed in helium-neon dimers using 
synchrotron radiation \cite{frank-group}. 

\begin{figure}[b]  
\vspace{-0.25cm}
\begin{center}
\includegraphics[width=0.33\textwidth]{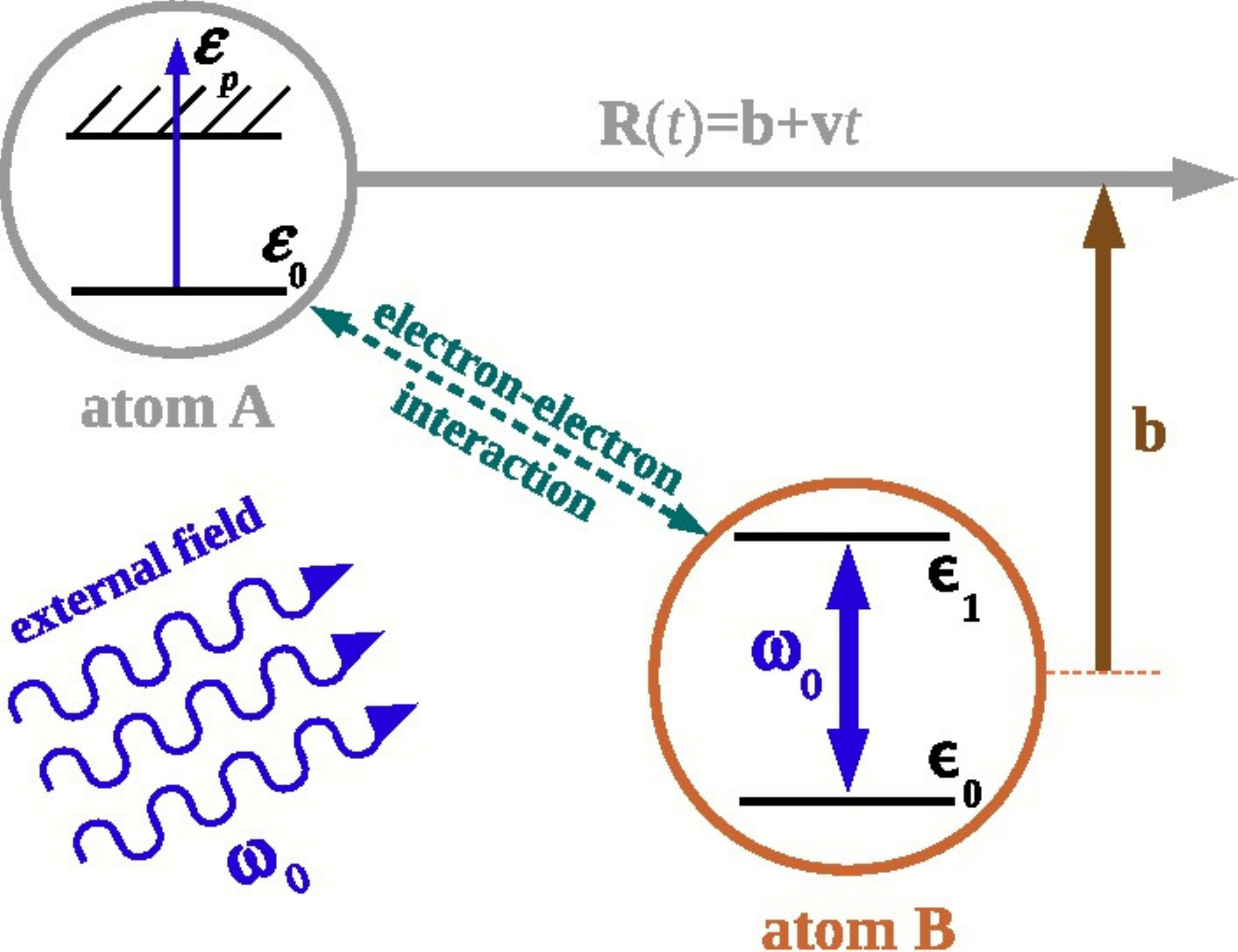}
\end{center}
\vspace{-0.5cm} 
\caption{ \footnotesize{ 
Scheme of photoionization in atomic collisions. }} 
\label{figure1}
\end{figure}

In this communication we study a dynamic variant of 
resonant two-center PI occurring 
in slow atomic collisions (see fig. \ref{figure1}). 
The average distance $R$ between colliding atoms 
is (many) orders of magnitude larger than in a bound system 
and the probability for the two-center PI scales as $R^{-6}$ 
\cite{we-2010}. Therefore, it might seem at first sight 
that in collisions the two-center ionization channel  
becomes already completely negligible. 
However, it turns out, quite unexpectedly,  
that it can dominate PI also in collisions. 

Atomic units are used throughout 
unless otherwise states. 

\section{General consideration} 

\subsection{Two-center ionization} 

Let us consider a collision between two atoms, $A$ and $B$,  
which are are initially (at $ t \to - \infty $) 
in their ground states, supposing    
that the binding energy in the ground state of $A$ 
is smaller than an excitation energy  
of a dipole allowed transition from the ground state  
in $B$. 
 
We shall assume the collision to be slow enough 
such that practically no excitation (or ionization) 
of the colliding partners is possible 
if $A$ and $B$ enter the collision being in their ground states. 
This is the case if $ \omega_{fi} a_0/v \gg 1$ 
(Massey adiabatic criterion, see e.g. \cite{massey-criterion}), 
where $\omega_{fi}$ and $a_0$  are typical transition frequency 
and linear size, respectively, of $A$ and/or $B$  
and $v$ is the collision velocity.  
 
However, if atom $B$ is coupled to 
an electromagnetic (EM) field resonant 
to a dipole transition between its ground 
and excited states,  
then the incident atom $A$ can be ionized 
by absorbing the excitation energy of $B$ 
via dynamic two-center electron correlations. 
 
We shall consider only very distant collisions, in which 
the interaction between $A$ and $B$ is quite weak and 
their nuclei move practically undeflected, 
and assume that, although the collision velocity is low, 
the relative motion of the nuclei   
can still be regarded as classical.    
In a reference frame, where atom $B$ is at rest and taken as 
the origin, atom $A$ moves along a classical straight-line 
trajectory ${\bf R}(t) = {\bf b} + {\bf v} t$, where 
${\bf b} = (b_x, b_y, 0)$ is the impact parameter and 
${\bf v} = (0, 0, v)$ the collision velocity. 
In this frame the collision is described 
by the Schr\"odinger equation 
\begin{eqnarray} 
i \frac{\partial \Psi(t)}{\partial t} = \hat{H}(t) \Psi(t), 
\label{e1}
\end{eqnarray}
where the total Hamiltonian is given by 
\begin{eqnarray} 
\hat{H}(t) = \hat{H}^A + \hat{H}^B + \hat{V}^{AB} + \hat{W}^A + \hat{W}^B.   
\label{e2}
\end{eqnarray}
Here, $\hat{H}^A$ ($ \hat{H}^B $) is the Hamiltonian of a free 
(non-interacting) atom $A$ ($B$), and 
\begin{eqnarray} 
\hat{V}^{AB} = \frac{ {\bf r} \cdot \boldsymbol{\xi} }{ R^3(t)} 
- 3  \frac{ \left( {\bf R}(t) \cdot {\bf r} \right) \, \,  
\left( {\bf R}(t) \cdot \boldsymbol{\xi} \right) }{ R^5(t)      }   
\label{e3}
\end{eqnarray}
is the interaction between $A$ and $B$, where 
${\bf r}$ ($ \boldsymbol{\xi} $) is the coordinate 
of the electron of $A$ ($B$) with respect to 
the nucleus of $A$ ($B$). Further, 
$\hat{W}^A$ ($\hat{W}^B$) is the interaction of $A$ ($B$)  
with the external EM field which will be  
taken as a classical linearly polarized field 
$ {\bf F}_0 \, \cos( \omega t - {\bf k} \cdot \boldsymbol{\xi} ) $ 
($ {\bf F} = {\bf F}_0 \, \cos( \omega t - {\bf k} \cdot ( {\bf R} + {\bf r} ) ) $), 
where $ {\bf F}_0 $ is the field strength, 
$ \omega $ the field frequency and ${\bf k}$ the wave vector 
($ {\bf F}_0 \cdot {\bf k} = 0$). 
The interactions $\hat{W}^A$ and $\hat{W}^B$ read  
\begin{eqnarray} 
\hat{W}^A &=& \frac{ {\bf A}({\bf r}, t) \cdot \hat{ {\bf p} }_{\bf r} }{ c } + 
\frac{ {\bf A}^2({\bf r}, t) }{2 c^2}
\nonumber \\ 
\hat{W}^B &=& \frac{ {\bf A}( \boldsymbol{\xi}, t) \cdot 
\hat{ {\bf p} }_{ \boldsymbol{\xi} } }{ c } + 
\frac{ {\bf A}^2(\boldsymbol{\xi}, t) }{ 2 c^2 },   
\label{e4}
\end{eqnarray}
where ${\bf A}( {\bf r}, t) = 
{\bf A}_0 \, \sin( \omega t - {\bf k} \cdot ( {\bf R} + {\bf r} ) ) $ 
(${\bf A}(\boldsymbol{\xi}, t) = 
{\bf A}_0 \, \sin( \omega t - {\bf k} \cdot \boldsymbol{\xi} ) $) 
with ${\bf A}_0 = - c {\bf F}_0/ \omega$ is the vector potential of the EM field 
at the position of the electron of atom $A$ ($B$) and 
$ \hat{ {\bf p} }_{\bf r} $ ($ \hat{ {\bf p} }_{ \boldsymbol{\xi} } $) is 
the momentum operator for the electron of atom $A$ ($B$). 
Below these interactions are taken in the dipole approximation: 
$ {\bf k} \cdot {\bf r} = 0$, $ {\bf k} \cdot \boldsymbol{\xi} = 0$.   

We first include the interaction between atom $B$ and 
the EM field by replacing  
the ground state $\phi_0$ (with an energy $\epsilon_0$) 
and the excited state $\phi_1$ (with an energy $\epsilon_1$)  
of non-interacting atom $B$ by its field-dressed bound states,   
\begin{eqnarray} 
\phi^{\pm}(t) &=& \alpha^{\pm}_0(t) \phi_0 + \alpha^{\pm}_1(t) \phi_1,      
\label{e5-general}
\end{eqnarray}
where $\alpha^{\pm}_0(t)$ and $\alpha^{\pm}_1(t)$ 
are time-dependent coefficients to be determined. 
We assume that the field is switched on adiabatically at $t \to - \infty$ 
and impose the boundary conditions $\phi^{+}(t \to - \infty) = 
\phi_0 \exp(- i \epsilon_0 t)$,  
$\phi^{-}(t \to - \infty) = \phi_1 \exp(- i \epsilon_1 t)$. 

Using the first order of perturbation theory 
in the interaction $\hat{W}^B$ 
we obtain  
\begin{eqnarray} 
\alpha^{+}_0(t) & = & \exp(- i \epsilon_0 t ) 
\nonumber \\ 
\alpha^{+}_1(t) & = & \frac{ W^B_{10} }{ \left(  \Delta  + i \Gamma_{rad}^{B}/2 \right) } 
\, \exp(- i ( \epsilon_0 + \omega ) t)
\label{e5-1order-plus}
\end{eqnarray}
and 
\begin{eqnarray} 
\alpha^{-}_0(t) & = & - \frac{ W^B_{01} }{ \left(  \Delta  + i \Gamma_{rad}^{B}/2 \right) } 
\, \exp(- i ( \epsilon_1 - \omega ) t)
\nonumber \\ 
\alpha^{-}_1(t) & = & \exp(- i \epsilon_1 t ),   
\label{e5-1order-minus}
\end{eqnarray}
where $ \Delta = \epsilon_0 + \omega - \epsilon_1$ is the 
detuning,  % from the resonance 
$\Gamma_{rad}^{B}$ 
the width of the excited state 
$\phi_1$ due to its spontaneous radiative decay  
and $W^B_{10} = 0.5 \, 
\langle \phi_1 | {\bf F}_0 \cdot \mbox{\boldmath{ $\xi$ }} | \phi_0 \rangle $ 
($W^B_{01} = (W^B_{10})^* $). 

If $ |W^B_{10}| > \Gamma_{rad}^{B} $ ($ |W^B_{10}| \gg \Gamma_{rad}^{B}$) 
the first order perturbation theory is no longer valid 
and instead we may use the rotating wave-approximation 
\cite{rotating-wave} to get    
\begin{eqnarray} 
\alpha^{+}_0(t) &=&  \sqrt{ \frac{ \Omega + | \Delta | }{ 2 \Omega } } 
\exp(- i ( \epsilon_{+} - \omega) t ) 
\nonumber \\ 
\alpha^{+}_1(t) &=& \frac{ 2 W^B_{10} }{ \sqrt{ 2 \Omega ( \Omega + | \Delta | ) } } 
\, \frac{ \Delta }{ |\Delta| } 
\, \exp(- i \epsilon_{+} t )    
\label{e5-rotat-wave-plus}
\end{eqnarray}
and 
\begin{eqnarray} 
\alpha^{-}_0(t) &=&  \sqrt{ \frac{\Omega - | \Delta | }{ 2 \Omega } } 
\exp(- i ( \epsilon_{-} - \omega) t ) 
\nonumber \\ 
\alpha^{-}_1(t) &=& - \frac{ 2 W^B_{10} }{ \sqrt{ 2 \Omega (\Omega - | \Delta | ) } } 
\, \frac{ \Delta }{ |\Delta| } \, \exp(- i \epsilon_{-} t ),     
\label{e5-rotat-wave-minus}
\end{eqnarray}
where $ \epsilon_{\pm} = \epsilon_1 + 0.5 ( | \Delta | \pm \Omega ) \, \Delta/| \Delta | $ 
are the quasi-energies of the field-dressed states.   
 
Using the states (\ref{e5-general}),  
the first order perturbation theory 
with respect to the interaction $\hat{V}^{AB}$, 
and keeping in mind that at $t \to -\infty$ both atoms 
were in the ground states we obtain that 
the two-center ionization amplitudes for atom $A$ reads  
\begin{eqnarray} 
a_{ 0 \to {\bf p}}^{2c,\pm} = i \int_{-\infty}^{+\infty} dt \exp( i( \varepsilon_p - \varepsilon_0) t)
\langle \psi_{\bf p} \phi^{\pm} | \hat{V}_{AB} | \psi_0 \phi^{+} \rangle  
\label{e8}
\end{eqnarray}
where $\psi_0$ with an energy $\varepsilon_0$ is 
the ground state of atom $A$ and 
$\psi_{\bf p}$ with an energy $ \varepsilon_p $ 
describes an electron emitted with 
an asymptotic momentum ${\bf p}$ (all the quantities 
refer to the rest frame of $A$) \cite{footnote1}.   

Performing the integration over time in (\ref{e8}) we obtain 
\begin{eqnarray} 
a_{ 0 \to {\bf p}}^{2c,\pm} &=& \frac{ 2 i \beta^{\pm} }{ v } s_p^{\pm} 
\left( K_1(s_p^{\pm}) \, 
\frac{ {\bf r}_{ {\bf p}, 0} \cdot \mbox{\boldmath{ $\xi$ }}_{0,1} - 
z_{ {\bf p}, 0} \, \xi_{z_{0,1}} }{ b^2 }  \right. 
\nonumber \\ 
&& \left. 
 + s_p^{\pm} K_0(s_p^{\pm}) \, \frac{ z_{ {\bf p}, 0} \, \xi_{z_{0,1}} }{ b^2 } \right. 
\nonumber \\ 
&& \left. - i \, q \, s_p^{\pm} K_1(s_p^{\pm}) \, 
\frac{ ( {\bf r}_{ {\bf p}, 0} \cdot {\bf b} ) \, \xi_{z_{0,1}} + 
( {\bf b} \cdot \mbox{\boldmath{ $\xi$ }}_{0,1} ) \, z_{ {\bf p}, 0} }{ b^3 } \right. 
\nonumber \\ 
&& \left. - s_p^{\pm} K_2(s_p^{\pm}) \, 
\frac{ ( {\bf r}_{ {\bf p}, 0} \cdot {\bf b} ) \, 
( {\bf b} \cdot \mbox{\boldmath{ $\xi$ }}_{0,1} ) }{ b^4 } \right).     
\label{e9}
\end{eqnarray}
Here, $ s_p^{+} = | \varepsilon_p - \varepsilon_0 - \omega | b /v  $, 
$ s_p^{-} = | \varepsilon_p - \varepsilon_0 - \omega - \Omega \, \Delta/| \Delta| | b /v  $, 
$ q = (\varepsilon_p - \varepsilon_0 - \omega)/| \varepsilon_p - \varepsilon_0 - \omega | $, 
$ {\bf r}_{ {\bf p}, 0} = \langle \psi_{\bf p} | {\bf r} | \psi_0 \rangle $,  
$ z_{ {\bf p}, 0} = \langle \psi_{\bf p} | z | \psi_0 \rangle $,
$ \mbox{\boldmath{ $\xi$ }}_{0,1} = \langle \phi_0 | \mbox{\boldmath{ $\xi$ }} | \phi_1 \rangle  $, $ \xi_{z_{0,1}} = \langle \phi_0 | \xi_{z}  | \phi_1 \rangle  $  
and $ K_n $ ($ n = 0, 1, 2$) are the modified Bessel functions \cite{IStegun}. 
In the  first order perturbation theory in the interaction 
$\hat{ W }^B $ 
$ \beta^{+} =  W^B_{10}/ \left(  \Delta  + i \Gamma_{rad}^{B}/2 \right) $ 
and $ \beta^{-} \approx 0 $. In the rotating-wave approximation  
$ \beta^{+} = W^B_{10}/\Omega $ and 
$ \beta^{-} = \sqrt{ (\Omega - | \Delta | )/(\Omega + | \Delta) |} \, W^B_{10}/\Omega $. 
 
The differential, $\frac{ d P^{2c}(b) }{ d^3 {\bf p}}$, 
and total, $ P^{2c}(b)$, ionization probabilities are given by   
\begin{eqnarray} 
\frac{ d P^{2c}(b) }{ d^3 {\bf p} } &=& \frac{ d P^{2c,+}(b) }{ d^3 {\bf p}} + 
\frac{ d P^{2c,-}(b) }{ d^3 {\bf p}} 
\label{spectra-1-new-1}
\end{eqnarray}   
with 
\begin{eqnarray} 
\frac{ d P^{2c,\pm}(b) }{ d^3 {\bf p} } &=& 
\frac{ 1 }{ 2 \, \pi  } \, 
\int_{ 0 }^{ 2 \pi } d \varphi_{\bf b} \, | a_{ 0 \to {\bf p}}^{2c,\pm} |^2,   
\label{spectra-1-new-2}
\end{eqnarray}   
where the integration runs over 
the azimuthal angle $ \varphi_{\bf b} $ 
of the impact parameter ${\bf b}$, and 
\begin{eqnarray} 
P^{2c}(b) = P^{2c,+}(b) + P^{2c,-}(b), 
\label{spectra-1-new-3}
\end{eqnarray}
with 
\begin{eqnarray} 
P^{2c,\pm}(b) &=&  \int d^3 {\bf p} \, 
\frac{ d P^{2c,\pm}(b) }{ d^3 {\bf p} }.        
\label{spectra-1-new-4}
\end{eqnarray}
The differential cross section, which describes 
the spectra of electrons emitted via the two-center channel 
in collisions with the impact parameters $b \geq b_{min} \gg 1$,  
read 
\begin{eqnarray} 
\frac{d \sigma^{2c} }{ d^3 {\bf p}} =  
\frac{d \sigma^{2c,+}}{ d^3 {\bf p}} + 
\frac{d \sigma^{2c,-}}{ d^3 {\bf p}}, 
\label{spectra-1}
\end{eqnarray}   
where 
\begin{eqnarray} 
\frac{d \sigma^{2c,\pm}}{ d^3 {\bf p}} &=& 2 \, \pi \, 
\int_{b_{min}}^{+\infty} db \, b \, \frac{ d P^{2c}(b) }{ d^3 {\bf p} }.   
\label{spectra-2}
\end{eqnarray}   
In (\ref{spectra-2}) the integration runs over the absolute value of 
the impact parameter. In particular, after some rather lengthy calculations 
we obtain that 
\begin{eqnarray} 
\frac{d \sigma^{2c,\pm}}{ d^3 {\bf p}} & = & \frac{ 1 }{ 2 \, \pi } \, 
| \beta^{\pm} |^2 \, | \xi_{01} |^2 \, 
\frac{ r^2_{p,0} }{ p^2 } \, \frac{ (s_p^{\pm})^2 \, b^2_{min} }{ v^4 }  \times 
\nonumber \\ 
&& \left(  (s_p^{\pm})^2 ( K^2_1(s_p^{\pm}) - K^2_0(s_p^{\pm}) ) \cos^2(\vartheta_p) 
\right. 
\nonumber \\ 
&& \left. + \left( s_p^{\pm} \, K_1(s_p^{\pm}) \, K_0(s_p^{\pm}) \right.  \right. 
\nonumber \\   
&& - \left. \left. \frac{ (s_p^{\pm})^2 }{ 2 } \, ( K^2_1(s_p^{\pm}) - K^2_0(s_p^{\pm}) ) \right)     
\sin^2(\vartheta_p) \right)    
\label{spectra-3}
\end{eqnarray}   
if the field is polarized along the $z$-direction (along the collision velocity) 
and 
\begin{eqnarray} 
\frac{d \sigma^{2c,\pm}}{ d^3 {\bf p}} & = & \frac{ 1 }{ 2 \, \pi } \, 
| \beta^{\pm} |^2 \, | \xi_{01} |^2 \, 
\frac{ r^2_{p,0}}{p^2} \,  
\frac{ (s_p^{\pm})^2 \, b^2_{min} }{ v^4 } \times 
\nonumber \\ 
&& \left(  s_p^{\pm} \, K_0(s_p^{\pm}) \, K_1(s_p^{\pm}) \right. 
\nonumber \\ 
&&\left. - \frac{1}{2} (s_p^{\pm})^2 ( K^2_1(s_p^{\pm}) - K^2_0(s_p^{\pm}) ) \cos^2(\vartheta_p) 
\right. 
\nonumber \\ 
&& \left. + \frac{1}{4} (s_p^{\pm})^2 ( K^2_1(s_p^{\pm}) - K^2_0(s_p^{\pm}) ) \sin^2(\vartheta_p) \right. 
\nonumber \\ 
&& \left. + \left( \frac{1}{2} \, K^2_1(s_p^{\pm}) + \frac{1}{8} \,(s_p^{\pm})^2 ( K^2_1(s_p^{\pm}) - K^2_0(s_p^{\pm}) ) \right) \right. 
\nonumber \\ 
&& \left. \times \sin^2(\vartheta_p) \right)  
\label{spectra-4}
\end{eqnarray}   
if the field is polarized along the $x$-direction. 
In (\ref{spectra-3})-(\ref{spectra-4}) 
$ r_{p,0} =  \int_0^{+\infty} dr \, r^3 \, u_{p1}(r) \, u_0(r) $ 
is the radial matrix element for transitions between the ground and 
continuum states of atom $A$ with $u_0$ and $u_{p,1}$ being 
their radial parts (the ground state of atom $A$ was assumed 
to be an $s$-state and $u_{p,1}$ denotes the continuum radial wave 
with the orbital quantum number $l = 1$). Further, 
$ \xi_{01} = \frac{ 1 }{ \sqrt{3} } \, \int_0^{+\infty} d\xi \, \xi^3 \, d_0(\xi) \, d_1(\xi)$ 
where $d_0$ and $d_1$ are the radial parts of the ground and excited states of atom $B$.  

The modified Bessel functions  
$K_n(x)$ ($n=0, 1, 2,..$) diverge at $ x \to 0$ 
and decrease exponentially at $x > 1$ \cite{IStegun}. 
Therefore, in distant low-velocity collisions 
($b \geq b_{min} \gg 1$, $v \ll 1$) the main contribution 
to the total cross section   
is given by a small interval of emission energies  
$ \delta \varepsilon_p \sim v/ b \ll 1 $ centred at 
$ \varepsilon_{p, r} = \varepsilon_0 + \omega $. 
If this interval is much less than a typical energy range 
$ \Delta \varepsilon_p $ in which the quantity $ r^2_{p,0} / p  $
substantially varies ($ \Delta \varepsilon_p \sim 10 $ eV for atoms and 
$ \Delta \varepsilon_p \sim 1 $ eV for negative ions), 
then  $ r^2_{p,0} / p  $ remains within $ \delta \varepsilon_p $ 
roughly a constant, $  r^2_{p,0} / p  \approx r^2_{p_r,0} / p_r  $ 
where $ p_r = \sqrt{2 ( \varepsilon_0 + \omega ) } $,   
and we obtain that     
\begin{eqnarray} 
P^{2c,\pm}(b) = \frac{ 3 \, \pi \, \alpha }{ 16 } |\beta^{\pm}|^2 \, | \xi_{01} |^2 
\frac{ r^2_{p_r,0} }{ p_r } \, \frac{ 1 }{ v \, b^5 },        
\label{e11}
\end{eqnarray}
\begin{eqnarray} 
P^{2c}(b) = \frac{ 3 \, \pi \, \alpha }{ 16 } 
\left( |\beta^{+}|^2 + |\beta^{-}|^2 \right) \, | \xi_{01} |^2 
\frac{ r^2_{p_r,0} }{ p_r } \, \frac{ 1 }{ v \, b^5 }          
\label{e11-new-1}
\end{eqnarray}
and 
\begin{eqnarray} 
\sigma^{2c,\pm}(b) &=& \frac{ \pi^2 \, \alpha }{ 8 \, v } 
| \beta^{\pm}|^2 \, | \xi_{01} |^2 
\frac{ r^2_{p_r,0} }{ p_r } \, \frac{ 1 }{ b_{min}^3 },         
\label{e11-new-2}
\end{eqnarray}
\begin{eqnarray} 
\sigma^{ 2c }(b) &=& \frac{ \pi^2 \, \alpha }{ 8 \, v } 
\left( |\beta^{+}|^2 + |\beta^{-}|^2 \right) \, | \xi_{01} |^2 
\frac{ r^2_{p_r,0} }{ p_r } \, \frac{ 1 }{ b_{min}^3 },           
\label{e11-new-3}
\end{eqnarray}
where $ \alpha = 1 $ ($ \alpha = \frac{ 3 }{ 2 } $) 
if the field is polarized along the $ z $-axis ($ x $- or $ y $-axis). 

Taking into account that atom $A$ moves in a gas of atoms $B$, 
the total ionization rate per unit of time 
via the two-center channel is given by 
\begin{eqnarray} 
\mathcal{K}^{2c} &=& \sigma^{2c} \, n_B \, v 
\nonumber \\ 
&=& \frac{ \pi^2 \, \alpha }{ 8 } \, 
\left( |\beta^{+}|^2 \, + \, |\beta^{-}|^2 \right) \, | \xi_{01} |^2 
\frac{ r^2_{p_r,0} }{ p_r } \, \frac{ n_B }{ b_{min}^3 },         
\label{e14}
\end{eqnarray}
where $n_B$ is the density of atoms $B$. 
The rate $\mathcal{K}^{2c}$, which turned out to be velocity independent, 
is proportional to $n_B$  
and by increasing it  
the two-center ionization process can be made 
more effective. However, there is a limitation 
on the upper boundary of $n_B$ set by the condition 
that the gas has to remain transparent for the EM field.  
At the resonance the excitation cross section  
reaches a large value: $ \sigma_{excit} = 3 \, \pi (c/\omega)^2 \gg \pi \, a_0^2 $. 
The mean free path $\lambda$ for the EM field 
in the gas of atoms $B$ can be estimated according to 
$ \lambda = 1/ ( n_B \sigma_{excit} ) $ and it has to be larger than 
the size $l_B$ of the gas target, $ \lambda > l_B$, 
in order that the target remains transparent for the EM field.    

\subsection{ Single-center ionization } 
The amplitude for the direct (single-center) ionization 
of atom $A$ is given by  
\begin{eqnarray} 
a_{ 0 \to {\bf p}}^{1c} &=& \frac{i}{2}  
\int_{-\infty}^{+\infty} dt \exp( i( \varepsilon_p - \varepsilon_0 - \omega) t) 
\langle \psi_{\bf p} \phi^{+}|{\bf F}_0 \cdot {\bf r} | \psi_0 \phi^{+} \rangle
\nonumber \\ 
&=& \pi i \, \langle \psi_{\bf p} |{\bf F}_0 \cdot {\bf r} | \psi_0  \rangle 
\delta\left( \varepsilon_p - \varepsilon_0 - \omega  \right).  
\label{e7}
\end{eqnarray}
This channel is described by the differential, $ d K^{1c}/d^3 {\bf p} $, 
and total, $ K^{1c}$, decay rates per unit of time which read   
\begin{eqnarray} 
\frac{d \mathcal{K}^{1c} }{ d^3 {\bf p}} = 
\frac{ 1 }{ 16 \pi }  \, \frac{ r^2_{p,0} }{ p^2 } 
\, f(\vartheta_{\bf p}, \varphi_{\bf p}) \, F_0^2 \, 
\delta( \varepsilon_p - \varepsilon_0 - \omega )  
\label{1-center-1}
\end{eqnarray}   
and    
\begin{eqnarray} 
\mathcal{K}^{1c} = \frac{ 1 }{ 12 }  \, \frac{ r^2_{p_r,0} }{ p_r } \, F_0^2,   
\label{1-center-2}
\end{eqnarray}   
where the value $p_r = \sqrt{ 2\, (\varepsilon_0 + \omega )}$ 
follows from the energy conservation 
expressed by the delta-function in (\ref{1-center-1}).  
The angular distribution is given by 
the function $f(\vartheta_{\bf p}, \varphi_{\bf p})$ depended   
on the polar, $\vartheta_{\bf p}$, and 
azimuthal, $\varphi_{\bf p}$,  emission angles of the electron. 
For instance,  $ f = \cos^2 \vartheta_{\bf p}$ if 
$ {\bf F}_0 = (0,0,F_0) $ and 
$ f = \sin^2 \vartheta_{\bf p} \, \cos^2 \varphi_{\bf p} $  
if ${\bf F}_0 = (F_0,0,0)$. 

\subsection{ Two-center versus single-center ionization } 
 
The competition between the two-center and single-center ionizations 
is natural to characterize by the ratio $\eta = \mathcal{K}^{2c}/\mathcal{K}^{1c}$. 
Using (\ref{1-center-2}) and (\ref{e14}) we obtain 
\begin{eqnarray} 
\eta  &=& \frac{ K_{2c} }{ K_{1c} }  
\nonumber \\ 
&=& \frac{ 3 \pi^2 \, \alpha }{ 2 } \, 
\left( |\beta^{+}|^2 \, + \, |\beta^{-}|^2 \right) \, | \xi_{01} |^2 
 \, \frac{ n_B }{ b_{min}^3 }.          
\label{e15}
\end{eqnarray}
Inserting into (\ref{e15}) expressions for $\beta^{\pm}$, derived within 
the first order % of perturbation theory 
in the interaction $ \hat{ W }^B$, results in 
\begin{eqnarray} 
\eta &=& \frac{ 3 \pi^2 \, \alpha }{ 8 } \, \frac{ n_B }{ b_{min}^3 } \, 
\frac{ | \xi_{01} |^4  }{ \Delta^2 + (\Gamma_{rad}^{B})^2/4    }.          
\label{e16}
\end{eqnarray}
Since $ \Gamma_{rad}^{B} = \frac{ 4 }{ 3 } \frac{ \omega^3 }{ c^3} \, | \xi_{01} |^2 $,  
at the resonance ($ \Delta = 0 $) the ratio becomes  
\begin{eqnarray} 
\eta &=& \frac{ 27 \pi^2 \, \alpha }{ 32 } \, \frac{ n_B }{ b_{min}^3 } \, 
\left( \frac{ c  }{ \omega } \right)^6.          
\label{e17}
\end{eqnarray}
In stronger fields, where the rotating-wave approximation should be used 
instead of the first-order perturbation theory, the ratio becomes smaller 
and decreases with increasing the field.  

\section{ Results and Discussion } 

Let us now apply formula (\ref{e17}) to the following collision systems: 

i) H(1s) (atom $A$, $ | \varepsilon_0 | \approx 13.6 $ eV) -- He(1s$^2$) (atom $B$); \\
ii) Li(1s$^2$ 2s) (atom $A$, $ | \varepsilon_0 | \approx 5.39 $ eV) -- H(1s) (atom $B$);\\ 
iii) K(4s) (atom $A$, $ | \varepsilon_0 | \approx 4.3 $ eV) -- Si (3p$^2$) (atom $B$); \\
iv) Li(2s) (atom $A$, $ | \varepsilon_0 | \approx 5.6 $ eV) -- Mg(3s) (atom $B$); \\
v) H$^-$(1s 1s') ('atom' $A$, $ | \varepsilon_0 | \approx 0.7 $ eV) -- Li(2s) (atom $B$); \\   
vi) H$^-$(1s 1s') ('atom' $A$, $ | \varepsilon_0 | \approx 0.7 $ eV) -- Rb(5s) 
(atom $B$).

i. H(1s) -- He(1s$^2$): Considering that the field is in resonance with 
the 1s$^2$ - 1s2p transition in helium ($\omega \approx 21$ eV) we obtain 
that at $b_{min} = 10$ a.u. $ \eta \geq 1$ provided   
$ n_B \geq 2.6 \times 10^{13}$ cm$^{-3}$. However, 
at $ n_B = 6.76 \times 10^{13}$ cm$^{-3}$ the mean free path $\lambda$ 
of the radiation in a gas of helium atoms would be merely $4.6 \times 10^{-3}$ cm. 
Thus, for this collision system a substantial enhancement of photo ionization
(by factor of $2$) from 
distant collisions due to the two-center channel would 
be possible only for very small-size helium gas targets.    
     
ii. Li(2s) -- H(1s): Assuming that the field is resonant to  
the 1s - 2p transition in hydrogen ($\omega \approx 10$ eV) 
$ \eta \geq 1$ with $b_{min} = 10$ a.u. is reached if  
$ n_B \geq 3 \times 10^{11}$ cm$^{-3}$. At the density 
$ n_B = 3 \times 10^{11}$ cm$^{-3}$  $\lambda \approx 0.09$ cm which means 
that the size of the target should not exceed $1$ mm in order that 
the two-center correlations in distant collisions 
yield a substantial contribution to PI. 

iii. K(4s) -- Si (3p$^2$): Considering that the field is in resonance with 
the 3p - 4s transition in silicon ($\omega \approx 4.9$ eV) we obtain  
that with $b_{min} = 10$ a.u. $ \eta \geq 1$ if   
$ n_B \geq 4.2 \times 10^{9}$ cm$^{-3}$.  
At $ n_B = 4.2 \times 10^{9}$ cm$^{-3}$ the mean free path $\lambda$ 
of the radiation in a gas of silicon atoms is about $ 1.6$ cm. 
Thus, for this collision system a substantial    
enhancement of PI from distant collisions due to the two-center channel 
would be possible for gas targets not exceeding $\sim 2$ cm.    
     
iv. Li(2s) -- Mg(3s): Assuming that the field is resonant to  
the 3s - 4p transition in magnesium ($\omega \approx 6.1$ eV) we get   
that $ \eta \geq 1$ at $b_{min} = 10$ a.u. 
provided $ n_B \geq 1.56 \times 10^{10}$ cm$^{-3}$. At the density 
$ n_B = 1.56 \times 10^{10}$ cm$^{-3}$  
$\lambda \approx 0.65 $ cm which means 
that the size of the target should not exceed $5$-$7$ mm 
in order that the two-center contribution from distant collisions 
doubles the ionization rate. 

v. H$^-$(1s 1s') -- Li(2s): Considering that the field is resonant 
to the 2s-2p transition in lithium ($\omega \approx 1.85$ eV) we obtain 
that $ \eta \geq 1$ at $b_{min} = 10$ a.u. is reached 
at $n_B \geq 1.2 \times 10^7$ cm$^{-3}$. 
At $n_B = 1.2 \times 10^7$ cm$^{-3}$ the mean free path of the radiation 
in lithium is already rather large, $\lambda \approx 77 $ cm. 
Since the typical size of targets in experiments with lasers is 
normally of the order of $1$ mm, 
one can increase the target density by about $800$ times  
which will reduce its size of transparency to the above $1$ mm. Then    
$ \eta \simeq 800 $ with $b_{min} = 10 $ a.u. and even with 
$b_{min}$ as large as  $ 50 $ a.u. one still obtains $ \eta \sim 6$-$7$.   
Thus, for H$^-$ -- Li system already very distant collisions may result 
in a strong enhancement of photo detachment from H$^-$ caused 
by the two-center ionization channel.  

vi. H$^-$(1s 1s') -- Rb(5s): Considering that the field is resonant 
to the 5s$_{1/2}$-5p$_{3/2}$ transition in rubidium ($\omega \approx 1.59$ eV) 
we obtain that at $b_{min} = 10$ a.u. $ \eta \geq 1$ provided  
$n_B \geq 4.88 \times 10^6$ cm$^{-3}$. 
At $n_B = 4.88 \times 10^6$ cm$^{-3}$ the mean free path of the radiation 
in rubidium is $\lambda \approx 140 $ cm. For rubidium targets 
with the size $ l_B \approx 1$ mm the transparency condition 
($\lambda \stackrel{<}{\sim} l_B $ cm) allows one to increase 
the target density up to  
$n_B \approx 1.4 \times 10^3 \times 4.88 \times 10^6 \approx  6.8 \times 10^9$ cm$^{-3}$ 
resulting in $ \eta \approx 1.4 \times 10^3$ with $b_{min} = 10 $ a.u. and even with 
$b_{min} = 50 $ a.u. the ratio is still quite large, $ \eta \approx 11$.   
Thus, for H$^-$ -- Rb collision system the effect is even larger than for 
H$^-$ -- Li one.    

Unlike the rates $\mathcal{K}^{1c}$ and $\mathcal{K}^{2c}$ 
their ratio $\eta$ does not depend on 
the transition matrix element of $A$. Therefore, 
we can apply (\ref{e16})-(\ref{e17}) 
also if 'atom' $A$ is in fact a molecule. In particular, 
for photo dissociation of I$_2$ ($ | \varepsilon_0 | \approx 1.57 $ eV) 
in collisions with Li ($\omega \approx 1.85$ eV) or Rb ($\omega \approx 1.59$ eV 
we obtain the same enhancements as for the H$^-$-- Li and H$^-$-- Rb systems.    
 
Since, according to our estimates, the inclusion 
of the contribution from collisions with $b < b_{min}$ 
(not taken into account here) strongly increases $\eta$,     
the effectiveness of the two-center channel seems to be 
really amazing. Indeed, let us put it into a perspective: 
in a gas of atoms $B$ with $n_B \sim 10^{10}$ cm$^{-3}$ 
the average distance $ R_{av} $  
between the atoms $A$ and $B$ ($ R_{av} \sim 1 / n_B^{1/3}$) 
is about $ 2.5 \times 10^{-4} $ cm $\sim 10^4$-$10^5 $ a.u.  
and nevertheless the two-center mechanism may still strongly dominate 
ionization of $A$. One of the reason for such a high effectiveness 
of this channel is that the effective mean 
distance $R_{eff}$ between the atoms 
in the collision, where the energy exchange between them 
is most likely to occur, is given by 
$ R_{eff} \sim \sqrt{ b_{min} / n_B^{1/3} } $ 
and turns out to be much smaller than $ R_{av} $.

\section{Conclusion} 

In conclusion, photoionization of atom $A$ in an external electromagnetic 
field can strongly increase if it moves with a low velocity 
in a gas of atoms $B$ which are in a dipole resonance with this field. 
This enhancement is caused by the transmission -- via the dynamic 
two-center electron correlations -- of photo-excitation 
energy from atom $B$, which acts as a very efficient antenna 
absorbing quanta from the field, to atom $A$ 
resulting in ionization of the latter.    

Two-center correlations are already 
known as an extremely efficient mechanism of 'communication'  
between parts of a bound system 
whose size $R$ is typically of the order of few or several Bohr radii. 
At $ R \gg 1 $ most of the two-center processes show 
the $R^{-6}$ scaling and the largest distances probed 
so far ($R\lesssim 10^2 $ a.u.)  
were in helium dimers \cite{He}.

In collisions the average distance between the atoms 
reaches tens of thousands of the Bohr radii. 
Nevertheless, the two-center correlations turn out 
to be quite effective also in collisions. 
This, in particular, suggests that a large variety 
of inter-atomic phenomena extensively investigated recently 
in bound systems, can play a role in 
collisions as well.

It is planned that a Rb gas target with up to $20 \%$ atoms 
transferred from the ground state ($5\, ^2S_{1/2}$  
to the excited $5\, ^2P_{3/2}$-state
by a weak resonant ($\omega \approx 1.5 $ eV) 
continuous laser field, which functions  
at the Institute of Modern Physics, will be combined in the near future 
with a beam of $\sim 100$ eV H$^{-}$ in order to       
perform an experiment on the two-center PI in slow atomic collisions.

\vspace{0.35cm}
{\bf Acknowledgement.} 

We thank B. Kullmann for his contribution 
at an early stage of this study 
and J. Yang for useful conversations.  
We acknowledge the support from the National Key 
Research and Development Program of China 
(Grant No. 2017YFA0402300),   
the German Research Foundation (DFG) 
(the projects MU 3149/4-1 and VO 1278/4-1),  
the program "One Hundred Talented People" 
of the Chinese Academy of Sciences 
and the CAS President's Fellowship Initiative.  
A. B. V. is grateful for 
the hospitality of the Institute of Modern Physics.


\begin{thebibliography}{99}

\bibitem{mcguire-book} 
J.H. McGuire, {\it Electron Correlation Dynamics in  
Atomic Collisions} (Cambridge University Press, 1997).  

\bibitem{abv-book-2008} A. B. Voitkiv and J. Ullrich, 
{\it Relativistic Collisions of Structured Atomic Particles} 
(Springer, Berlin, 2008).  

\bibitem{Smirnov} B. M. Smirnov, Sov. Phys. Usp. {\bf 24}(4), 251 (1981).

\bibitem{Weidemueller} S. Giovanazzi, A. Gorlitz, and T. Pfau, Phys. Rev.
Lett. {\bf 89}, 130401 (2002); T. Amthor, M. Reetz-Lamour,
S. Westermann, J. Denskat, and M. Weidem\"uller, Phys.
Rev. Lett. {\bf 98}, 023004 (2007).

\bibitem{Forster} T. F\"orster, Ann. Physik \textbf{437}, 55 (1948). 
 
\bibitem{Suhai} S. Suhai, Phys. Rev. B \textbf{51}, 16553 (1995).

\bibitem{solids}
T. P. Devereaux and R. Hackl, Rev. Mod. Phys. 79, 175 (2007)

\bibitem{icd-all} L. S. Cederbaum, J. Zobeley, and F. Tarantelli, 
Phys. Rev. Lett. {\bf 79}, 4778 (1997); R. Santra and L. S. Cederbaum, Phys.
Rep. {\bf 368}, 1 (2002); 
V. Averbukh et al., J. Electron Spectrosc. Relat. Phenom. {\bf 183}, 36 (2011); 
U. Hergenhahn, ibid. {\bf 184}, 78 (2011); 
T. Jahnke, J. Phys. {\bf B 48}, 082001 (2015). 

\bibitem{clusters} S. Marburger \textit{et al.}, Phys. Rev. Lett. \textbf{90}, 203401 (2003).

\bibitem{resonantICD} T. Aoto \textit{et al.}, Phys. Rev. Lett. \textbf{97}, 243401 (2006).

\bibitem{He}
N. Sisourat \textit{et al.}, Nature Phys. 6, 508 (2010);
T. Havermeier \textit{et al.}, Phys. Rev. Lett. \textbf{104}, 133401 (2010).

\bibitem{ICDexpH2O} T. Jahnke \textit{et al.}, Nature Phys. \textbf{6}, 139 (2010);
M. Mucke \textit{et al.}, \textit{ibid.} \textbf{6}, 143 (2010).

%\bibitem{matthew} J. Matthew and Y. Komninos, Surf. Sci. \textbf{53}, 716 (1975).

%\bibitem{kay} A. Kay et al., Science \textbf{281}, 679 (1998). %;

%\bibitem{gokhberg} K. Gokhberg et al., Europhys. Lett. \textbf{72}, 228 (2005). 

\bibitem{2CDR} C. M\"uller, A. B. Voitkiv, J. R. Lopez-Urrutia, and Z. Harman,
Phys. Rev. Lett. {\bf 104}, 233202 (2010); 
A. B. Voitkiv and B. Najjari, Phys. Rev. {bf A 82}, 052708 (2010).  

\bibitem{our-new-paper} A. Eckey et al, Phys. Rev. {\bf A 98} 012710 (2018). 

\bibitem{SPDI}
R. D\"orner \textit{et al.}, Phys. Rev. Lett. \textbf{77}, 1024 (1996).

\bibitem{Rzazewski}
K. Rzazewski and J. H. Eberly, Phys. Rev. Lett. \textbf{47}, 408 (1981);
G. S. Agarwal \textit{et al.}, \textit{ibid.} \textbf{48}, 1164 (1982).

\bibitem{NSDI} B. Walker \textit{et al.}, Phys. Rev. Lett. \textbf{73}, 1227 (1994). 

\bibitem{we-2010} B. Najjari, A. B. Voitkiv and C. M\"uller, 
Phys. Rev. Lett. {\bf 105} 153002 (2010); 
A. B. Voitkiv and B. Najjari, Phys. Rev. {\bf A 82}, 052708 (2010); {\bf 84} 013415 (2011).

\bibitem{frank-group} F. Trinter, J. B. Williams, M. Weller, M. Waitz, M. Pitzer, 
J. Voigtsberger, C. Schober, G. Kastirke, C. M\"uller, C. Goihl,
P. Burzynski, F. Wiegandt, R. Wallauer, A. Kalinin, L. P. H.
Schmidt, M. S. Sch\"offler, Y. -C. Chiang, K. Gokhberg, T. Jahnke,
and R. Dorner, Phys. Rev. Lett. {\bf 111}, 233004 (2013); 
A. Mhamdi et al., Phys. Rev. {\bf A 97}, 053407 (2018).

\bibitem{massey-criterion} 
H. S. W. Massey, Rep. Prog. Phys. 12, 248 (1949); 
D. R. Bates, Phys. Reports 35, 305 (1978); 
J. B. Delos, Rev. Mod. Phys. {\bf 53} 287 (1981). 
 
\bibitem{rotating-wave} P. L. Knight and P. W. Milloni, Phys. Rep. {\bf 66} 
21 (1980); M. V. Fedorov and A. E. Kazakov, 
Prog. Quant. Electr. {\bf 13} 1 (1989). 

\bibitem{footnote1} We neglect the Doppler shift and the coupling to 
the scalar potential which appears in a moving reference frame 
\cite{abv-galilei-gauge} that is justified due to the low collision velocity.   

\bibitem{abv-galilei-gauge} A. B. Voitkiv, J. Phys. {\bf B 39 } 4275 (2006). 

\bibitem{IStegun} M. Abramowitz and I. Stegun,
{\it Handbook of Mathematical Functions}
(Dover Publications, Inc., New York, 1965). 
  
\end{thebibliography}
\end{document}